\documentclass[%
 reprint,
 amsmath,amssymb,
aps,
superscriptaddress]{revtex4-2}

\usepackage{graphicx}
\usepackage{dcolumn}
\usepackage{bm}
\usepackage{xcolor}

\begin{document}

\preprint{APS/123-QED}

\title{CMOS on-chip thermometry at deep cryogenic temperatures}

\author{Grayson M.~Noah}
\email{These authors contributed equally to this work}
\affiliation{Quantum Motion, 9 Sterling Way, London, N7
9HJ, United Kingdom}
\author{Thomas Swift}
\email{These authors contributed equally to this work}
\affiliation{Quantum Motion, 9 Sterling Way, London, N7
9HJ, United Kingdom}
\affiliation{London Centre for Nanotechnology, UCL,
London, WC1H 0AH, United Kingdom}
\author{Mathieu de
Kruijf}
\email{These authors contributed equally to this work}
\affiliation{Quantum Motion, 9 Sterling Way, London, N7
9HJ, United Kingdom}
\affiliation{London Centre for Nanotechnology, UCL,
London, WC1H 0AH, United Kingdom}
\author{Alberto~Gomez-Saiz}
\affiliation{Quantum Motion, 9 Sterling Way, London, N7
9HJ, United Kingdom}
\author{John J. L. Morton}
\email{These authors jointly lead this work}
\affiliation{Quantum Motion, 9 Sterling Way, London, N7
9HJ, United Kingdom}
\affiliation{London Centre for Nanotechnology, UCL,
London, WC1H 0AH, United Kingdom}
\author{M.~Fernando~Gonzalez-Zalba}
\email{These authors jointly lead this work}
\affiliation{Quantum Motion, 9 Sterling Way, London, N7
9HJ, United Kingdom}

\date{\today}

\begin{abstract}
Accurate on-chip temperature sensing is critical for the optimal performance of modern CMOS integrated circuits (ICs), to understand and monitor localized heating around the chip during operation. The development of quantum computers has stimulated much interest in ICs operating a deep cryogenic temperatures (typically 0.01 - 4 K), in which the reduced thermal conductivity of silicon and silicon oxide, and the limited cooling power budgets make local on-chip temperature sensing even more important. Here, we report four different methods for on-chip temperature measurements native to complementary metal-oxide-semiconductor (CMOS) industrial fabrication processes. These include secondary and primary thermometry methods and cover conventional thermometry structures used at room temperature as well as methods exploiting phenomena which emerge at cryogenic temperatures, such as superconductivity and Coulomb blockade. We benchmark the sensitivity of the methods as a function of temperature and use them to measure local excess temperature produced by on-chip heating elements. Our results demonstrate thermometry methods that may be readily integrated in CMOS chips with operation from the milliKelivin range to room temperature.

\end{abstract}

\maketitle 

Heat is a common byproduct of information-processing technologies~\cite{Landauer1961}. Inefficiencies in the information conversion process have led to power dissipation densities approaching those inside a nuclear reactor core, compromising the chip integrity due to the elevated temperatures and as a result steering the development of modern computing technologies~\cite{Pop2010,Ferain2011}. Quantum information processing technologies, although theoretically dissipationless, are also subject to these power dissipation challenges, particularly taking into account that useful quantum computers will require large numbers of qubits operating dynamically at microwave frequencies~\cite{Fowler2012, OGorman2017, Reilly2019, Bravyi2022}. Moreover, the `classical' digital and analogue electronics which are used for the control, addressing and read-out of quantum processors dissipate power like any conventional circuit. 
The majority of leading quantum processor platforms demonstrate optimum behaviour at deep cryogenic temperatures (0.01 to 4~K), including superconducting~\cite{cryoCMOSbardin, supercqubits}, spin~\cite{cryosi, sixqubitsi}, trapped ion~\cite{cryooinons,trappedion2} and photonic~\cite{cryophotonics} qubits, and hence the goal of more tightly integrating the control electronics with quantum processors has motivated developing cryogenic ICs~\cite{Hornibrook2015,Charbon2016,Patra2018,gonzalezzalba2021}. 

At these deep cryogenic temperatures, cooling power is orders of magnitude smaller~\cite{Pobell1996}, while the thermal conductivity of silicon and silicon oxide substantially reduce when compared to room temperature~\cite{Slack1964}. Therefore, cryogenic ICs require power management solutions and on-chip thermometry methods capable of monitoring hotspots across the chip operating at temperatures down to 10s of milliKelvin (mK). 

A number of on-chip thermometry techniques have been adopted across the semiconductor industry. Silicon bandgap temperature sensing utilizes the temperature dependence of the voltage across a P-N junction in a bipolar junction transistor (BJT) at a given applied current --- this is typically used in the $T=30-500$~K temperature range; however, with a combination of advanced diode design and low bias current, improved sensitivity for temperatures approaching 1.5~K has been demonstrated~\cite{Courts2002,Shwarts2008,COURTS2016,Xue2021}. 
Silicided polycrystalline silicon (polysilicon) structures are also used as on-chip temperature sensors based on their temperature coefficient of resistance (TCR), which can be constant down to 50{K}, but vanishes around 30~K~\cite{Hart2021} (depending on the silicide material). As silicided polysilicon is also often used in the gate stack of complimentary metal-oxide-semiconductor (CMOS) processes, this allows for temperature measurement directly above a field effect transistor (FET) in a technique known as gate resistance thermometry (GRT)~\cite{Pavlidis2017}. This high degree of localisation is especially important in ultra-thin-body (UTB) technologies such as silicon-on-insulator (SOI) and FinFETs, where oxide barriers below and/or on the sides of the UTB trap heat within the channel.  
GRT has been used to show that the temperature of a current-carrying transistor can be more than 10~K higher than that of other transistors on the same chip~\cite{Triantopoulos2019}.

\begin{figure}[ht!]
\includegraphics[width=\columnwidth]{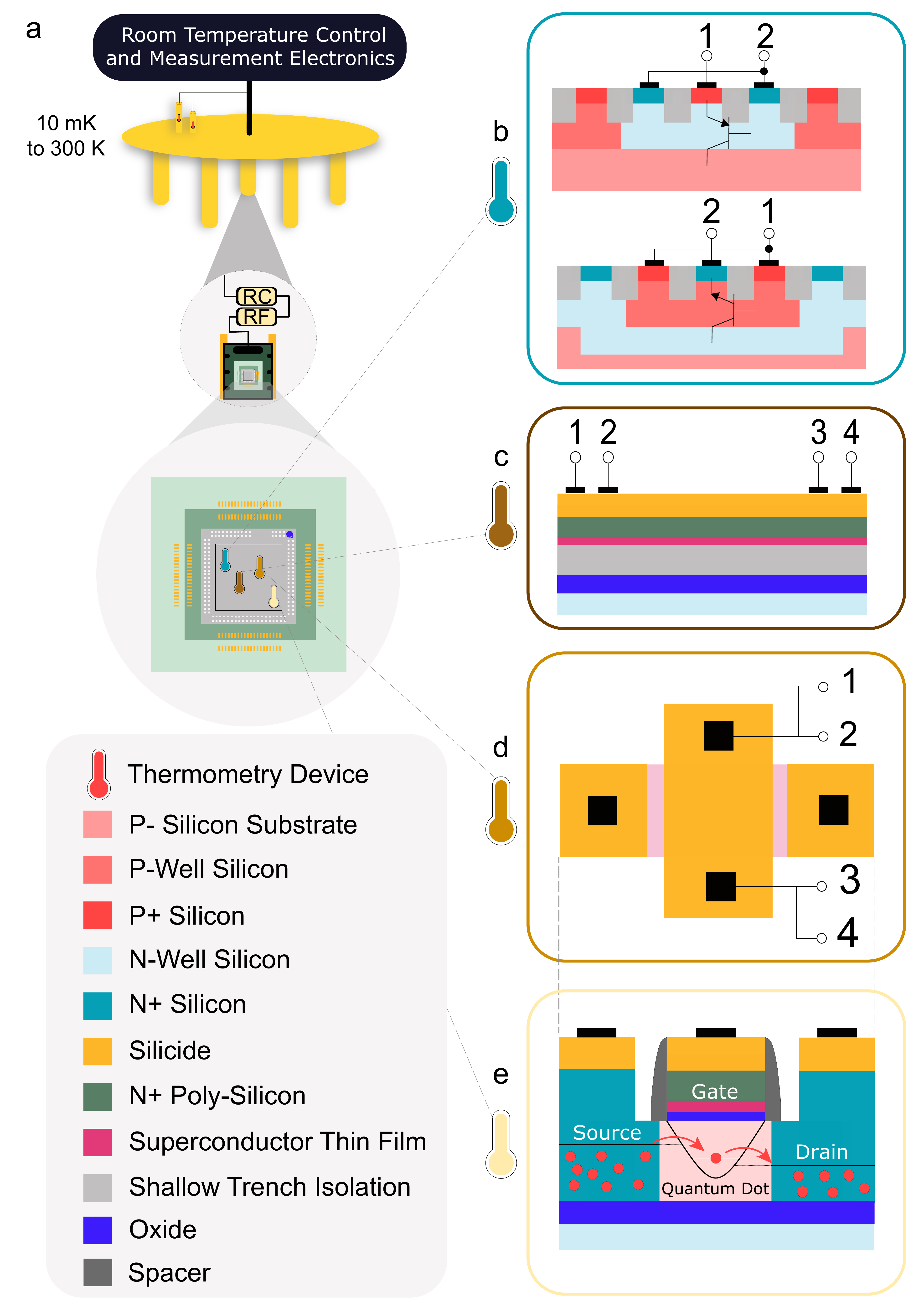}
\caption{\label{fig:1_Measurement_Setup} \textbf{(a)} Diagram of cryogenic measurement setup. DC signals are carried from room-temperature electronics to the mixing chamber using DC PhBr looms thermalized at each stage of a dilution refrigerator and then RF and DC filtered. The 3~mm x 3~mm chip is glued to a PCB. The legend in the bottom left gives the materials in (b)-(e). \textbf{(b)} NPN and PNP diode structures \textbf{(c)} Silicided polysilicon resistor structure with contacts on silicide layer allowing 4-point measurement \textbf{(d)} Field-effect transistor with gate stack similar to \textbf{c}, measurement contacts are separated away from device. \textbf{(e)} Quantum dot transistor with overlay showing the energy level structure that results in sequential single-electron tunnelling.}
\end{figure}

Here, we investigate four solutions for on-chip thermometry for CMOS ICs which can be operated down to the mK temperature range relevant for many quantum technologies. We explore to what extent existing approaches can be adapted to operate in such conditions, as well as studying new opportunities for thermometry which emerge at these low temperatures. In particular, we study how diode thermometry based on silicon P-N junctions and GRT perform even at temperatures below 1 K, well outside of their usual operating ranges, and we identify a method of superconducting phase transition thermometry based on the critical current of the superconducting thin films which are present as part of the CMOS process. Each of the above methods provides a secondary thermometer which requires calibration; we also present a primary method, quantum dot thermometry, based on tunneling conductance measurements through discrete energy levels of a quantum dot formed using the CMOS process. 
We first introduce the four different thermometry methods in Sec.~\ref{sec:1}. Next, we present their operating principle and calibration procedure in Sec.~\ref{sec:2} followed by a comparative benchmark in terms of sensitivity in Sec.~\ref{sec:3}. Finally, we utilize the four methods to measure excess on-chip heating in Sec.~\ref{sec:4} and discuss the results in Sec.~\ref{sec:5}.

\section{\label{sec:1}On-chip temperature-sensing techniques}

We first present the devices utilized for on-chip thermometry as well as describe the measurement setup (see  Fig.~\ref{fig:1_Measurement_Setup}a). All devices are contained within a 3~mm x 3~mm chip fabricated using a UTB SOI process from an industrial foundry. Most devices are placed in arrays, with each individual device addressable using on-chip multiplexing.
We power up the chip using 0.8 and 1.8~V supplies resulting in static power dissipation during operation. We bond the chip to a printed circuit board (PCB) that sits inside a sample puck using silicon-doped aluminum bondwires. The puck is physically connected to the mixing chamber (MXC) of a dilution refrigerator with a base temperature of 20~mK. At the MXC stage, two thermometers are placed approximately 40~cm from the sample: a BlueFors ruthenium oxide and a PT100 platinum. These thermometers measure the MXC temperature ($T_\text{MXC})$ and can be used for temperature calibration (see Methods \ref{SuppA}), but as we discuss later, the accuracy of the MXC thermometers to chip temperature is compromised by the physical separation, highlighting the need for on-chip temperature readings.   

\begin{figure*}[ht!]
\includegraphics[width=\textwidth]{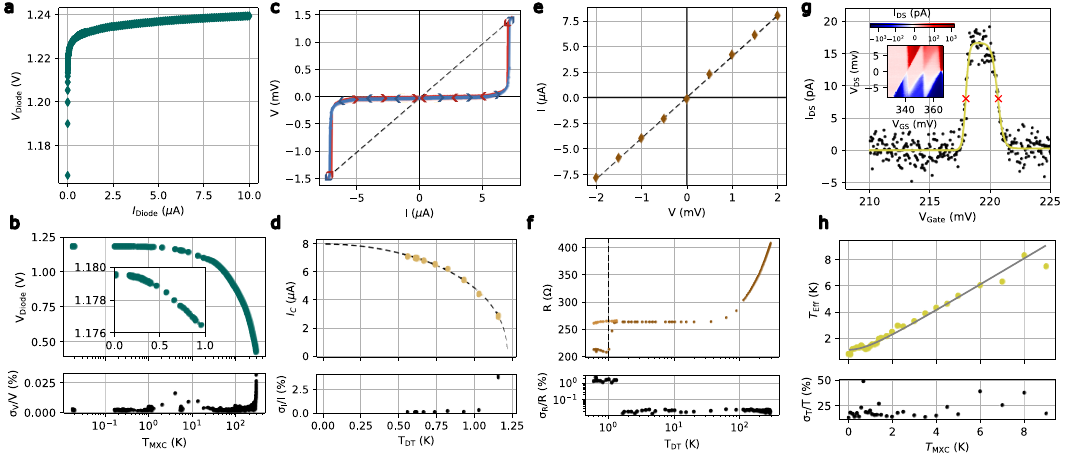}
\caption{\label{fig:2_Measurement_Procedures} \textbf{(a)} PNP diode IV sweep at 20mK. \textbf{(b)} $V_{\text{Diode}}$ (top) and associated measurement precision (bottom) at $I_\text{diode}=1$~nA vs MXC temperature for an NPN diode. \textbf{(c)} IV curve for 4-point measurement of superconducting thin film in the polysilicon resistor. \textbf{(d)} Critical current vs diode temperature fitted with adapted Bardeen formula (Eq.~\ref{eq:Bardeen}). (Bottom) Precision of the measurement vs diode temperature. \textbf{(e)} IV curve of quasi-4-point measurement of the gate of an FET with linear regression fit. \textbf{(f)} Resistance (top) and precision (bottom) vs diode temperature for the gate showing a superconducting transition at 1.2~K. \textbf{(g)} IV sweep of single-electron transistor at constant source-drain bias ($V_\text{ds}=2$~mV) showing the fitted top hat lineshape (yellow line) taken at a back-gate voltage of 2~V. The red crosses mark the $V_\text{0,s(d)}$ points that enable extracting the lever arm, $\alpha_\text{g}=V_\text{ds}/\left[V_\text{0d}-V_\text{0s}\right]$. (Inset) Coulomb blockade map taken at a back gate voltage of 0~V. \textbf{(h)} (Top) Extracted temperature from the Fermi fits $T_\text{eff}$ vs $T_\text{MXC}$ (yellow dots) and fit $T_\text{eff}=\sqrt{T_\text{MXC}^2+T_0^2}$ (black line) showing the one-to-one dependence and saturating temperature $T_0$. (Bottom) Precision of the measurement vs $T_\text{MXC}$.}
\end{figure*}

\textbf{Diode thermometry (DT)}: We start with silicon vertical diode structures, specifically the two varieties PNP (top) and NPN (bottom) shown in Fig.~\ref{fig:1_Measurement_Setup}b. Both structures use shallow trench isolation (STI) to define the junction. The PNP diode is forward-biased by applying a positive voltage to a highly-doped P-type contact region relative to the voltage of the N-well directly beneath it, whereas the NPN structure is designed to operate with the opposite bias configuration. 
We use a constant current to forward-bias the diode, and measure the voltage between contacts 1 and 2 as a function of temperature, $V_\text{Diode}(T)$. Unless otherwise stated, we use the NPN diode in the rest of the Article.

\textbf{Superconducting phase transition thermometry (SPTT)}: Next, in Fig.~\ref{fig:1_Measurement_Setup}c, we show the layered structure of a polysilicon resistor commonly used in CMOS processes due to the stability of its resistance value at high temperatures. This example consists of a silicided highly-doped N-type polysilicon layer on top of a thin adhesion layer of superonducting thin film on SiO$_2$. As discussed later, the superconducting layer transitions at a critical temperature $T_{\text{C}}\sim 1.2$~K. We probe the superconducting nature of the film using a 4-point measurement, enabling investigation of the temperature-dependence of the critical current, $I_\text{C}(T)$.

\textbf{Gate resistance thermometry (GRT)}: A layered structure similar to the polysilicon resistor also forms the gate stack of an FET (length, $L=150$~nm, and width, $W=2$~$\mu$m) as seen in Fig.~\ref{fig:1_Measurement_Setup}d. We extract the temperature-dependent resistance of the gate using a 4-point measurement. For thermometry, we use the normal-state resistance $R_\text{gate}(T)$ above $T_\text{C}$ and the critical current of the superconducting thin film below. This method provides the most local temperature reading of the corresponding FET.    

\textbf{Quantum dot thermometry (QDT)}: Finally, we consider the FET in Fig.~\ref{fig:1_Measurement_Setup}e. Due to its small dimensions ($L=28$~nm and $W=80$~nm), the transistor can be utilized to trap individual electrons in an electrostatically defined quantum dot (QD) formed in the silicon channel directly under the gate electrode~\cite{Hofheinz2006,Yang2020b}. When operated at deep cryogenic temperatures near threshold, the silicon below the gate spacers is highly resistive due to the gradual drop of doping density from the ohmic to the channel, effectively forming tunnel barriers between the source/drain reservoirs and the QD~\cite{Voisin2014}. The lineshape of the drain-source current versus gate-source voltage can be utilized as a local temperature sensor~\cite{Zwanenburg2013, Maradan2014}.      

We now describe the calibration sequence for the different thermometers. We first calibrate DT to the MXC since the diodes can be measured without powering the auxiliary circuitry, minimising the on-chip static power dissipation and hence allowing DT to closely track the MXC temperature down to 20~mK (see the calibration protocol in Methods \ref{SuppA}).  The remaining sensors are all measured through multiplexers which require powering the additional circuitry leading to a minimum temperature of $T_\text{on}\approx 600$~mK. This figure does not represent a hard physical limit but is specific to the auxiliary circuit used here. Therefore using $T_\text{MXC}$ to directly calibrate the remaining sensors may lead to inaccurate readings. To increase the accuracy in the calibration, we use the diode to calibrate the remaining secondary thermometers, SPTT and GRT. Finally, QDT can be self calibrated.

\section{\label{sec:2}Operation principle}

\textbf{DT:} Figure \ref{fig:2_Measurement_Procedures}a shows an example of a PNP diode IV curve measured at $T_\text{MXC}=20$~mK, showing a typical exponential dependence of the current ($I_\text{Diode}$) with applied voltage across the junction ($V_\text{Diode}$). The threshold voltage of the junction is temperature-dependent which effectively results in a change in $V_\text{Diode}$ as the temperature is changed. Figure~\ref{fig:2_Measurement_Procedures}b shows this temperature dependence for a bias current $I_\text{Diode}=1$~nA. We select this low bias current as it minimizes self-heating while maintaining the precision of the measurement ($\sigma_V/V_\text{Diode}$ where $\sigma_V$ is the standard deviation) below 0.02\% from $20$~mK to room temperature, see bottom of Fig.~\ref{fig:2_Measurement_Procedures}b. We note that significant temperature sensitivity is maintained down to 100~mK, outside the specified operating range of any commercial silicon diode thermometers~\cite{Shwarts2008,COURTS2016}. This result may be due to the comparatively lower $I_\text{Diode}$ used in this experiment or other factors such as doping concentrations and geometry. We calibrate the DT response to a weighted combination of the readings from the MXC ruthenium oxide and the platinum thermometer assuming perfect thermalization (see Methods~\ref{SuppA}).       

\textbf{SPTT:} We probe the superconducting thin films using a 4-point measurement to eliminate the influence of series resistances including both that of the multiplexer and vertical contact resistances through the metal and polysilicon layers. Figure~\ref{fig:2_Measurement_Procedures}c shows a typical hysteretic IV curve with a slightly larger upsweep (switching) critical current (blue line) and a slightly smaller downsweep (retrapping) critical current (red line). The switching critical current ($I_\text{C}$) is an intrinsic property of the superconductor which is stochastic in nature and has a temperature-dependence following the Bardeen formula~\cite{Bardeen1962}. The retrapping current arises due to resistive heating~\cite{Tinkham2003}, which makes it less sensitive for use in thermometry, as the temperature of the thin film is already elevated above the phonon bath. Figure~\ref{fig:2_Measurement_Procedures}d shows the measured $I_\text{C}$ vs $T_\text{DT}$ (varied by changing the MXC temperature), fitted using an adapted form of the Bardeen formula (dashed line) which has been used to accurately fit the dependence of the critical current with temperature in the vicinity of the critical temperature~\cite{Kuznetsov_2017}, 

\begin{equation}
\label{eq:Bardeen}
    I_{\text{C}}(T) =  I_{\text{C}}(T=0) \left [ 1- \left(\frac{T}{T_{\text{C}}} \right )^2\right ]^i.
\end{equation}

Here we use the critical temperature $T_{\text{C}}$, the critical current at zero temperature $I_{\text{C}}(0)$, and the exponent $i$ as fitting parameters . We find $I_{\text{C}}(0)$ = 7.95 $\pm$ 0.06~$\mu$A, $T_{\text{C}}(0)$ = 1.21 $\pm$ 0.01~K and $i=0.44$ $\pm$ 0.02. The form of $I_\text{C}(T)$ means it is most sensitive to $T$ as it approaches $T_{\text{C}}$ and by applying an out-of-plane magnetic field, the sensor's point of maximum sensitivity can be tuned (see Methods~\ref{SuppB}). We calibrate $I_\text{C}$ to the diode temperature, making the SPTT a secondary thermometer. We also note that we use the SPTT in a complementary way to standard SPTT where the sharp slope of the $R-T$ curve at the superconducting-normal transition is used for sensitive thermometry over a narrow range of temperatures~\cite{Luo2014}. Finally, in the bottom panel, we show the precision as a function of $T_\text{DT}$ which varies from 0.1\% to 4\% as the temperature approaches $T_\text{C}$.     

\textbf{GRT:} We measure the resistance across the silicided gate structure of an FET by performing a 4-point IV measurement and using linear regression to fit the data as shown in Fig.~\ref{fig:2_Measurement_Procedures}e. The resistance decreases with temperature as shown in Fig.~\ref{fig:2_Measurement_Procedures}f, which is the expected result for a metal~\cite{Pobell1996}. The TCR is constant down to 50~K and then goes to zero at 30~K below which the resistance does not change. However, the layer of superconducting thin film transitions at $T_\text{C}$ which is shown by a sharp drop in resistance below 1.2~K (the residual resistance is due to the metal-silicide contact which is part of the 4-wire measurement for the FET gate). The diamond points denote the normal resistance for a bias current above $I_\text{C}$, demonstrating that this is a superconducting transition rather than a normal resistivity change of the material. We calibrate the thermometer to the diode, making GRT a secondary thermometer. We note the precision in the resistance measurement $\sigma_R/R$ is of the order of 0.01\% in the normal state and deteriorates to approximately 1\% in the superconducting region.

\textbf{QDT:} The technique uses IV measurements through a discrete state in a QD to infer the electronic temperature of the charge reservoirs. In the inset of Fig.~\ref{fig:2_Measurement_Procedures}g, we first confirm the presence of a QD in the channel of the FET by monitoring the drain-source current, $I_\text{ds}$, as a function of the gate-source $V_\text{gs}$ and drain-source $V_\text{ds}$ voltages. We observe the characteristic diamond-shape regions of low conductance, i.e. Coulomb diamonds~\cite{Kouwenhoven2001}, indicative of charge quantization. We perform a gate voltage sweep ($V_\text{ds}=2$~mV) through the first Coulomb oscillation to reveal a top-hat lineshape in the measured current. This lineshape is characteristic of electronic transport through a discrete quantum state~\cite{Ahmed2018}. The data can be fit to a sequential single-electron tunneling expression,

\begin{equation}
    I_{\mathrm{ds}} = e\frac{\Gamma_{\mathrm{s}} \Gamma_{\mathrm{d}}}{\Gamma_{\mathrm{s}}+\Gamma_{\mathrm{d}}}, 
    \label{eq:tophat}
\end{equation}

\noindent in which $e$ is the electronic charge, $\Gamma_\text{s(d)}=\Gamma_\text{0,s(d)}f(\varepsilon_\text{s(d)})$ is the source(drain) reservoir-to-QD tunnel rate ~\cite{Houten1992,Maradan2014} and $f_\text{s(d)}(\varepsilon)$ is the Fermi distribution evaluated at the source(drain). Here, $\Gamma_\text{0,s(d)}$ is the maximum source(drain) tunnel rate and $\varepsilon_\text{s(d)}=-e\alpha_\text{g}\left[V_\text{gs(d)}-V_\text{0,s(d)}\right]$ is the energy detuning between the source(drain) Fermi level and the QD electrochemical level, which align at $V_\text{gs(d)}=V_\text{0,s(d)}$. Finally, the lever arm, $\alpha_\text{g}$, the ratio between the gate and total capacitance of the QD, can be extracted from the Coulomb diamond measurement~\cite{Ihn2010} giving this sensor a self-calibrating nature. We obtain $\alpha_\text{g}=0.91\pm 0.10$, see Fig.~\ref{fig:2_Measurement_Procedures}g. Since the current through the device is related to temperature by the Fermi function, which is a known physical law, the sensor is a primary thermometer. We utilize the falling edge of the lineshape for temperature extraction since it avoids the impact of excited states in the measurement bias window~\cite{Ahmed2018b}. In Fig.~\ref{fig:2_Measurement_Procedures}h, we plot the extracted temperature $T_\text{QDT}$ as a function of the MXC temperature. From $6$~K down to approximately 1.5~K, we observe a one-to-one linear dependence between the QDT and the MXC. Below 1.5~K, the QDT reading starts to saturate and the thermometer becomes inaccurate. At $T_\text{MXC}=1.12\pm0.05$~K, the reading plateaus. The reason for this deviation at the lower-temperature end is associated with an elevated charge noise in the sample that broadens the top-hat lineshape beyond the thermal limit. The precision of the measurement is predominantly below 25\% for the low temperature range but increases as the temperature exceeds ~6~K, i.e. the Fermi level broadening approaches the drain-source excitation ($3.5k_\text{B}T_\text{MXC}\approx eV_\text{ds}$). 

\begin{figure*}[ht!]
\centering
\includegraphics[width=\textwidth]{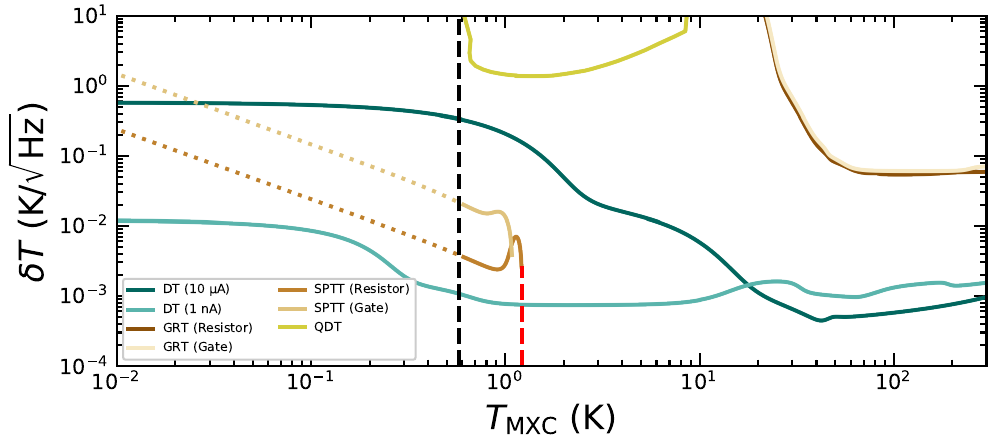}
\caption{Plots of the bandwidth-normalized sensitivity vs temperature (MXC for DT and QDT and diode temperature for SPTT and GRT). Diode thermometry for $I_\text{Diode}=1$~nA (light blue) and $I_\text{Diode}=10$~$\mu$A (dark green). Superconducting phase transition thermometry below $T_\text{C}$ and normal resistance thermometry above $T_\text{C}$ (brown). Gate resistance thermometry below and above $T_\text{C}$ (light yellow). Quantum dot thermometry (yellow). The red dashed line represents the superconducting transition temperature and the black dashed line is the minimum operation temperature of the chip, $T_\text{on}$.}
\label{fig:3_sensitivity}
\end{figure*}

\section{\label{sec:3}Sensitivity Benchmark}

Next, we proceed to benchmark the different thermometry methods. To compare thermometers, we calculate the bandwidth-normalized sensitivity,

\begin{equation}
    \delta T = \frac{1}{\sqrt{\text{BW}}}\frac{\sigma_{\mathrm{A}}}{\partial A/\partial T},
    \label{eq:sensitivity}
\end{equation}

\noindent which corresponds to the minimum resolvable temperature change in a measurement of bandwidth BW. Here $A$ and $\sigma_A$ are the average and the standard deviation of the observable, respectively. Using the results presented in Fig. \ref{fig:2_Measurement_Procedures}, we extract the sensitivity of each method and present the data in Fig.~\ref{fig:3_sensitivity} as a function of temperature. For a discussion of the measurement bandwidth of the different methods, see Methods~\ref{Suppc}. 

We first discuss the sensitivity of DT. The low-current-bias diode ($I_\text{Diode}=1$~nA, light blue trace) shows a sensitivity O(1 mK/$\sqrt{\text{Hz}}$) from room temperature down to 300~mK before deteriorating due to self-heating of the sensor. However, a substantial sensitivity remains even down to the lowest measured temperatures. For comparison, we show a high-current-bias result (industry-standard current of $I_\text{Diode}=10$~$\mu$A, dark green trace) where the sensitivity start to deteriorates below 15~K. For the other measurement techniques, the chip needs to be powered on, setting a lower bound on the temperature that can be measured at $T_\text{on}\approx 600$~mK (black dashed line).

Next, we discuss the SPTT sensitivity for both the standalone polysilicon resistor and the FET gate material. Above $T_\text{C}$ (red dashed line), the technique is insensitive to temperature. Just below $T_\text{C}$, the sensitivity presents a minimum due to the sharp derivative of $I_\text{C}$ near $T_\text{C}$. However, as the temperature is reduced, the high error in the determination of the critical current close to $T_\text{C}$ and the reduced derivative results in a local maximum. As the temperature is reduced further, $\sigma_{I}$ is reduced and the sensitivity improves reaching an optimal sensitivity point. Finally, as the temperature lowers well below $T_\text{C}$, the sensitivity deteriorates due to the flattening of the $I_\text{C}-T$ dependence. For temperatures below $T_\text{on}$, we use the fit to the modified Bardeen formula to extend the expected sensitivity which we indicate by the dashed extensions of the data. The sensitivity of SPTT in the FET gate material is generally worse than that of that of the standalone polysilicon resistor due to its lower resistance (due to the physical dimensions) and inclusion of contact resistances in its 4-wire measurements increasing the error in detection of the switching current transition. We also include the sensitivity of the normal-state resistances of each device, i.e. GRT (or regular resistance thermometry in the case of the standalone polysilicon resistor). Above 50~K, the sensitivity is stable at O(0.1~K/$\sqrt{\text{Hz}}$) and deteriorates sharply below due to the saturation of the resistance by scattering crystal lattice imperfections~\cite{Pobell1996}. The sensitivity dependence with temperature is similar for both the polysilicon resistor and the FET gate material. 

Finally, we discuss QDT. We note that since QDT is a primary thermometry technique, the observable is temperature and hence Eq.~\ref{eq:sensitivity} reduces to $\delta T= \sigma_T/\sqrt{\text{BW}}$. The sensitivity is therefore determined by the standard deviation for a given measurement bandwidth. At high temperatures, $T\gtrsim 6$~K, the sensor's sensitivity is limited by the standard deviation of the measurement. Particularly, when the Fermi width approaches the applied source-drain bias ($3.5k_\text{B}T\approx eV_\text{ds}$) the current passing through the device decreases, reducing the precision as well as the accuracy of the measurement. At intermediate temperatures ($1.5-6$~K) the sensor presents its region of optimal sensitivity O(1~K/$\sqrt{\text{Hz}}$). At even lower temperatures, the sensitivity is limited by a combination of the elevated $T_\text{on}$ and charge noise in the device. Furthermore, we must note that below 1.5~K, the sensor accuracy deteriorates, as described in Sec.~\ref{sec:2}. The comparatively poor sensitivity of QDT with respect to the other methods is a consequence of the low bandwidth of the measurement (common to primary thermometers that require a functional fit to extract temperature~\cite{Pekola1994,Spietz2003, Iftikhar2016}), needing to acquire a full IV trace with relatively small currents (10s of picoAmperes). 

Having discussed the operation principles and benchmarked the sensitivity of the four different methods, we present a summary of the different specifications in Table~\ref{Table1}. Overall, we find that DT provides the most sensitive and wide-ranging method for on-chip thermometry, particularly when biased using a low current (1~nA). One of the reasons for this superior sensitivity is the high measurement bandwidth as only a single data point is utilized to extract the temperature reading. Bandwidth is reduced for other thermometry methods which require a series of data points to be acquired for valid fit of the data; this is particularly intensive for SPTT and QDT, both of which require a very fine step size in their IV curves to extract a precise temperature reading. With regards to the temperature range of operation, the physical mechanism for DT is robust in the range of temperatures studied. On the other hand, SPTT and QDT are limited in range due to the physical nature of their mechanisms. The normal-superconducting transition sets an upper limit for the SPTT while the Coulomb blockade effect does so for the QDT. In the case of QDT, charge noise is a limiting factor at the low-temperature end. This limitation is not an intrinsic effect to QDT but rather to the particular technology of the sample. In the case of GRT, in its resistive form, a low-temperature limit exists ($\approx 30$~K) due to impurity scattering. Finally, we note that although QDT presents an inferior sensitivity to DT (and SPTT below $T_\text{C}$), its primary nature allows for self-calibration without the need of a separate thermometer.    

\begin{table*}
\label{Table1}
\centering
\begin{tabular}{|l|r|r|r|r|}
\hline
\textbf{Specification} & \textbf{DRT} & \textbf{GRT} & \textbf{SPTT} & \textbf{QDT} \\ \hline\hline
\textbf{\begin{tabular}[c]{@{}l@{}}Minimum\\ Temperature (K)\end{tabular}}  & 0.13 & 30 & 0.6*  & 1.12** \\ \hline
\textbf{\begin{tabular}[c]{@{}l@{}}Maximum\\ Temperature (K)\end{tabular}} & 300+ & 300+ & 1.2 & 6*** \\ \hline
\textbf{Sensitivity (K/$\sqrt{\text{Hz}}$)}                                                                 & 0.001-0.01 & 0.1-1 & 0.05-0.2 & 1-10 \\ \hline
\textbf{Precision (\%)}                                                                    & (2-7)$\cdot 10^{-3}$ & (2-3.5)$\cdot 10^{-2}$ & 0.07-4 & 10-25 \\ \hline
\textbf{Type}                                                      & Secondary  &  Secondary & Secondary & Primary \\ \hline
\textbf{Integration Difficulty}                                                      & Low & Low & Moderate & Very High \\ \hline
\end{tabular}
\caption{\label{Table1}Benchmark table. Comparison between the four different methods in terms of minimum and maximum temperature, sensitivity, precision, type of thermometer and integration capability within an integrated circuit (including readout). Precision is defined as $\sigma_A/A$. Integration difficulty here represents a relative measure of the level of design challenge and quantity of independent circuit functionalities required (e.g. DAC, ADC, etc.) to be integrated on the same chip. We consider an integrated sensor a circuit without any external analog electronic instruments such that a digital code representing temperature could be returned by the chip. * Power up limited. **Charge noise limited. ***Coulomb blockade limited. }
\end{table*}

\begin{figure*}[ht!]
\includegraphics[width=\textwidth]{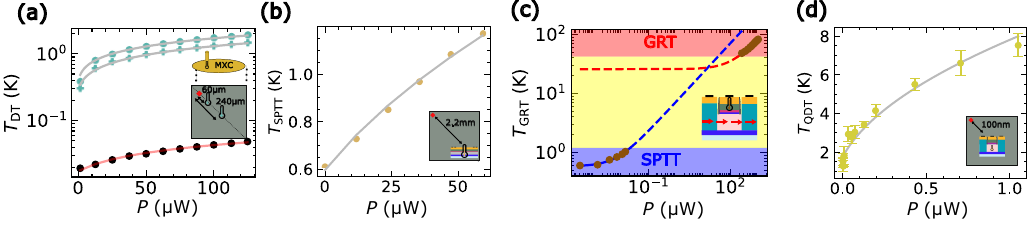}
\caption{\label{fig:4_remote_heating} On-chip temperature versus dissipated power measurements. \textbf{(a)} Measured using DT and heating with NPN diodes at 60~$\mu$m and 40~$\mu$m (blue dots). Corresponding MXC temperature (black dots). \textbf{(b)} Measured using SPTT and heating with a diode at 2.2~mm. \textbf{(c)} Self-heating measurements for FET devices using GRT (high-temperature region) and SPTT (low-temperature region). The blue fit extrapolates from the low-temperature (SPTT) points only, whereas the red fit extrapolates from the high-temperature (GRT) points only. As such, the red fit's low-temperature tapering in the yellow region should not be taken to represent a real effect since the sensor cannot measure low enough temperatures to determine its true $T_\text{0}$, but the blue fit's predictions at high power are more reasonable, as we expect the square-root trend to continue upwards. \textbf{(d)} Measured using QDT and heating with an FET at 100~nm.}
\end{figure*}

\section{\label{sec:4}Measuring on-chip heating}

Having calibrated the different thermometers and described their specifications, we now measure locally the effects of on-chip power dissipation. To heat up the chip, rather than using the MXC heaters, we utilize on-chip diodes and FETs located at different distances from the actual thermometers (see Fig.~\ref{fig:supp_chip_locs} in Methods~\ref{SuppE}).

We first present local heating measurements using DT (see Fig.~\ref{fig:4_remote_heating}a). We heat the calibrated NP diode thermometer by driving a current through two other diodes located at distances of 60~$\mu$m and 240~$\mu$m. We observe a sublinear increase of the diode temperature with heater power, $P$, and a more pronounced increase for the 60~$\mu$m case, as expected. We observe a similar sublinear temperature increase with power for the SPTT, GRT and QDT as we shall discuss later (Fig.~\ref{fig:4_remote_heating}b-d). Additionally, in Fig.~\ref{fig:4_remote_heating}a, we plot the MXC temperature at each power level (black dots) and plot it along with the on-chip temperature demonstrating the disparity between the two readings and underlining the need for on-chip thermometry. In this example, $T_\text{DT}$ reaches a temperature higher than 1.8~K for the 60~$\mu$m case, whereas monitoring only the MXC (which remains below 50~mK throughout) would provide no indication of such on-chip temperature rise.

To obtain a quantitative understanding of the thermal mechanisms playing a role in the steady-state on-chip temperature, we develop a model that takes into account the power dissipation, cooling power and thermal resistances present within the system. These include the quadratic dependence of the cooling power of a dilution refrigerator with temperature~\cite{Richardson1988, Iftikhar2016} as well as the dependencies of the thermal conductivity of the metals ($\sim T$) and insulating materials ($\sim T^3 $) involved~\cite{Duthil2014} (see Methods~\ref{SuppF}). This leads to a relation between the sensor temperature, $T$, and the heater power, $P$, as well as the background power sources, $P_\text{s}$, that elevate the chip temperature above $T_\text{MXC}$ when no power is applied,

\begin{equation}
\label{eq:Tt4a}
    T^4 = T^3\alpha\sqrt{P+P_{\text{s}}}+\beta_\text{h}P+\beta_\text{s}P_{\text{s}}.
\end{equation}

The term including the proportionality constant $\alpha$ represents the joint effects of the varying metal thermal conductivity and MXC cooling power with temperature. Overall, the term presents a square root dependence of the sensor temperature with applied power. On the other hand, the terms preceded by the fitting constants $\beta_\text{h}$ and $\beta_\text{s}$ represent, respectively, the effect of the heater power and the background power on the sensor temperature via the insulating materials. In this case, a fourth root dependence is observed. 

Coming back to DT, we perform a fit to Eq.~\ref{eq:Tt4a} considering that the background sources of power are comparatively small ($P_\text{s}\rightarrow 0$~W) since the digital and auxiliary circuity in this experiment are powered off. We find that both the square and fourth root terms are important in performing an accurate fit of the reading, particularly at the lowest temperatures. However, we find that in the case of SPTT, GRT and QDT, a fit to the simplified version of Eq.~\ref{eq:Tt4a}, without the contribution of the insulating elements,

\begin{equation}
    T = \sqrt{\alpha^2 P+T_\text{0}^2},
    \label{eq:power}
\end{equation}

\noindent is sufficient to perform accurate fits of the different thermometer readings (see Methods~\ref{SuppF}). Here, we have defined $T_\text{0}=\alpha^2P_\text{s}$ as the base temperature of the sensor. We recall that, in these cases, the support circuitry is powered up, elevating the base temperature of the chip to $T_\text{on}\approx$600~mK. Our results suggest that at temperatures close to the MXC base temperature, the insulating materials play an important role in thermalizing the sensor. However, as the temperature increases, the dominant cooling mechanisms are the combined effect of the increased thermal conductivity of the metals, further facilitating thermalization, and the additional MXC cooling power. 

We now look more closely at SPTT. A PNP heater diode is approximately 2.2~mm from the polysilicon resistor structure, allowing a similar experiment to be conducted using the critical current of the superconducting structure across different power levels as shown in Fig.~\ref{fig:4_remote_heating}b. This method can only be used while the chip is powered up and hence the temperature range is limited to values between $T_\text{on}$ and $T_\text{C}$. A fit to Eq.~\ref{eq:power}, reveals an $\alpha^\text{SPTT}= 130\pm20$~KW$^{-1/2}$ and $T_\text{0}^\text{SPTT}=597\pm 13$~mK, the latter in close agreement with $T_\text{on}$.

When performing GRT, the FET remains fully operational and, given the close distance from the channel of just a few nanometers, the experiment can effectively be considered a self-heating test. To vary the power dissipated in the channel, we sweep $V_\text{ds}$ at fixed $V_\text{gs}$. For small power dissipation, we use the superconductivity of the gate stack, effectively an SPTT measurement which for the purpose of differentiation we refer as superconducting GRT (SGRT). We show the resulting points below 1.2~K in Fig.~\ref{fig:4_remote_heating}c. From the fit, we find $\alpha^\text{SGRT}= (7.2\pm1.5)\cdot 10^3$ KW$^{-1/2}$, and $T_\text{0}^\text{SGRT}=583\pm12$~mK due to on-chip power dissipation of the auxiliary circuitry, as discussed above. For higher powers, the gate temperature increases to the point where the normal resistance of the silicide layer becomes temperature-sensitive, as shown in the points above 30~K. Here, we find $\alpha^\text{GRT}= (2.46\pm 0.02)\cdot 10^3$~KW$^{-1/2}$, and $T_\text{0}^\text{GRT}=25.7\pm 1.3$~K. The lower $\alpha$, compared to that of the superconducting layer, is at least partially a consequence of the larger separation between the silicide layer and the channel. The extracted $T_\text{0}^\text{GRT}$ is within the range in which TCR tends to zero.
The technique shows sensitivity to the most extreme levels of on-chip heating allowing self-heating characterisation across 6 orders of magnitude in dissipated power with a dead zone in between. More particularly, we see that the FET gate temperature exceeds 1~K with just a few tens of nanowatts and reaches nearly 100~K with 1~mW of power dissipation.

Finally, we discuss QDT. To heat up the QDT, we use another identical FET placed parallel to the QDT at an edge-to-edge distance between channels of 100~nm. We drive a current through the FET by using a constant $V_\text{gs}$ and variable $V_\text{ds}$ resulting in a static power dissipation of up to 1~$\mu$W. Again, we observe a sublinear increase in temperature as a function of power well-modelled by Eq.~\ref{eq:power}. We find $\alpha^\text{QDT}=(8\pm2)\cdot 10^3$~KW$^{-1/2}$ indicating a higher degree of sensitivity to power dissipation than SGRT despite the higher locality of the latter. The reason for the larger $\alpha$ could be associated to the larger power density produced in the FET channel heating the QD (recall $W\times L=80\times28$~nm) that in the FET used in the GRT experiments ($W\times L=2000\times150$~nm).  

Overall, the measurements presented in this section demonstrate the capability of the four different sensors to measure local excess heating well above the temperatures detected by the thermometers at the MXC. These results highlight the major importance of on-chip thermometry when assessing local temperature hotspots since they provide a much more accurate reading of the true chip thermal environment and shed light on the cooling mechanism playing a role.

\section{\label{sec:5}Conclusions}

The performance of solid-state quantum computers and cryo-electronics microcircuits is closely linked to the temperature of their environment. Understanding hence what the local temperature is in dynamically operated circuits and architectures is of primary importance. Here, we have introduced and benchmarked four different methods for local on-chip thermometry native to CMOS technology such as they could be seemingly integrated in scaled-up circuits. We find that DT, particularly when biased with a low current (1~nA) is the most sensitive method when compared to GRT, SPTT and QDT for the whole temperature range studied (20~mK to 300~K). Particularly, the low current bias technique enables sensitive thermometry well below 1.5~K, the common lower limit for commercial cryogenic diode thermometers. We envision that the sensitivity of the DT technique could be improved further by operating the sensor in conjunction with fast readout techniques~\cite{vigneau2022}. Such an approach could surpass state-of-the-art thermal sensitivity figures~\cite{Chawner2021, Blanchet2022} while remaining compatible with industry-standard fabrication processes and may enable the study of thermal dynamics at deep cryogenic temperatures in the microsecond timescale.

\section{Methods}

\subsection{\label{SuppA}Calibration of the chip temperature}

\begin{figure}[ht!]
\includegraphics[width=\columnwidth]{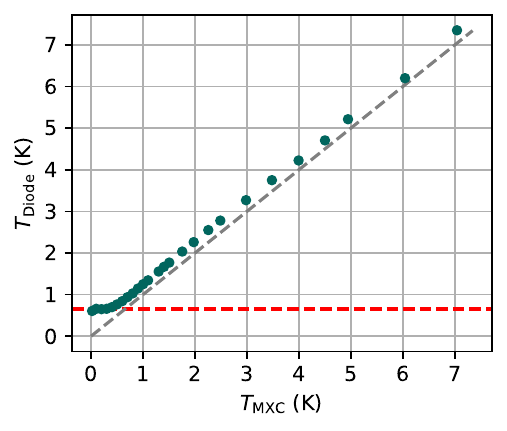}
\caption{\label{fig:supp_diode_mxc}Plot of measured PNP diode temperature vs MXC chamber temperature with the chip powered on. Grey dashed line shows the path the measurement should track for perfect thermalization and no power dissipation (y=x). The red dashed line is the point at which the diode temperature plateaus ($\text{T}_{\text{on}}\sim$~600~mK) which represents the minimum operation temperature of the chip when powered.}
\end{figure}

Apart from the diodes, all the structures used in this study required powering up the additional digital and analogue circuitry to enable access through the cryogenic multiplexers. The power used by the support circuitry consumes 4.3~$\mu$W and raises the on-chip temperature above that of the MXC temperature. To extract the base chip temperature when the support electronics were active, $T_\text{on}$, we used DT. 

In Fig. \ref{fig:2_Measurement_Procedures}a, we described the calibration procedure of the DT to the MXC thermometer which we now use to calibrate $T_\text{on}$. Specifically, we use the diode at 1~nA bias current and measure the temperature, $T_\text{DT}$ as a function of $T_\text{MXC}$ (see Fig.~\ref{fig:supp_diode_mxc}). We fit $T_\text{DT}=\sqrt{T_\text{MXC}^2+T_\text{on}^2}$ and extract $T_\text{on}$ =631 $\pm$ 24~mK. We conclude that the static power dissipation raises the chip temperature above the reading of the MXC thermometers highlighting the necessity of placing thermometers on chip. We make the assumption that all on-chip structures are at the same temperature when the digital and auxiliary circuitry is powered up and no other heat sources are applied. Due to the distribution of devices and auxiliary circuitry across the chip, this may not be the case, leading to partial errors. Nevertheless, it provides a much more accurate picture of the chip temperature than direct readings from the MXC thermometer. A more advanced version may benefit from creating an array of thermometers to extract a thermal image of the powered-on chip. 

\subsection{\label{SuppB} Magnetic field dependence of SPTT}

As discussed in Sec. \ref{sec:2}, the critical current of a superconductor is a temperature-dependent quantity that can be used in thermometry, as given by Eq.\ref{eq:Bardeen}. However, in Fig. \ref{fig:3_sensitivity}, we saw that the sensitivity of this technique is best for $T\lesssim T_{\text{C}}$. To increase the range of high sensitivity, the dependence of the critical current with magnetic field can be exploited. More particularly, both $T_\text{C}$ and $I_\text{C}$ are magnetic-field-dependent properties and are given by Eq.~\ref{eq:TC_B}\&\ref{eq:IC_B}.

\begin{equation}
\label{eq:TC_B}
    T_\text{C}(B_{\perp}) = \text{T}_{\text{C}}(B_{\perp}=0) \sqrt{1- \frac{B_{\perp}}{\text{B}^{\perp}_{\text{C}}(T=0)}}
\end{equation}

\begin{equation}
\label{eq:IC_B}
    I_\text{C}(B_{\perp}) = \frac{\text{I}_{\text{C}}(B_{\perp}=0)}{1+\frac{B_{\perp}}{B_0}} 
\end{equation}

These equations can be combined to fit critical current data across a wide range of temperatures and magnetic fields. The equations use as fitting parameters: the critical out-of-plane magnetic field at zero temperature $B_\text{C}^\perp(T=0)$; the critical current and temperature at zero magnetic field, $I_\text{C}(B_\perp=0)$ and $T_\text{C}(B_\perp=0)$; and  $B_0$ which is a macroscopic materials parameter based on a Kim-Type fit~\cite{Kim1962}. Based on the equations, we see that an increase in the out-of-plane magnetic field leads to a decrease in the critical current and the critical temperature. This decrease in critical temperature shifts the region of optimal sensitivity of the technique to lower temperatures, showing how the optimal sensitivity of the SPTT can be tuned using external magnetic fields.

\subsection{\label{Suppc}Measurement bandwidth}

In Sec.~\ref{sec:3}, we use the bandwidth-normalized sensitivity to benchmark the different thermometers. This ensures that the link between standard deviation of the measurement and integration time is taken into account. In this Appendix, we describe the details of the measurement bandwidth of each method used to quantify the sensitivity. 

For DT, we use an integration time of 20 ms per point resulting in a measurement bandwidth of 50~Hz. For SPTT and GRT (and resistance measurements in general), we use 20~ms integration time for each point and take a 5~ms settling time between points. The bandwidth for these techniques is therefore $BW = \frac{1}{N * 25~ms}$, where $N$ is the number of points (9 for resistance measurements and variable for critical current measurements depending on step size and critical current magnitude). The determination of the resistance measurements from a set of points rather than single point measurements ensures that temperature variations in the chip leakage currents do not affect the resistance reading. Finally, for the Coulomb blockade measurements the bandwidth is 0.1~Hz, as each peak trace acquisition takes 10~s before it is then fit to extract the temperature.

\subsection{\label{SuppD}Critical Current Distribution}

\begin{figure}[h!]
\includegraphics[width=\columnwidth]{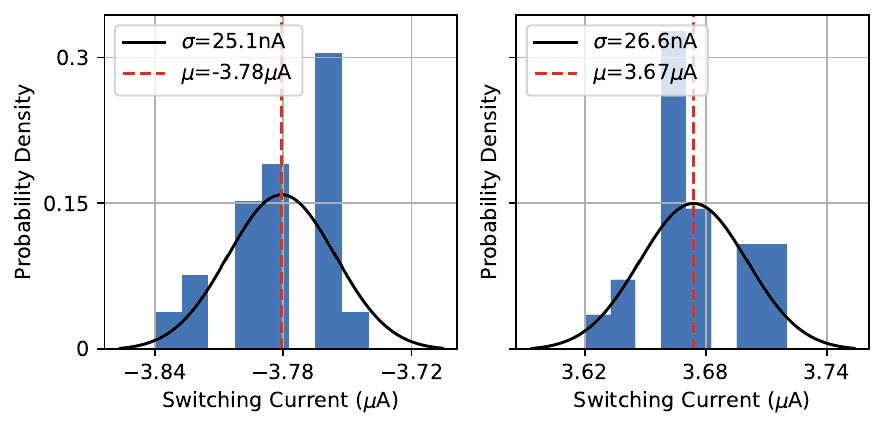}
\caption{\label{fig:supp_Ic} Probability density of the critical current as a function of the switching current for both the positive and negative current flow directions. }
\end{figure}

In Fig.~\ref{fig:supp_Ic}, we plot the critical current distribution from ten IV sweeps. We fit the measurement outcomes to a Gaussian distribution to determine an equivalent standard deviation of the critical current value. The distribution of measured critical currents represents the combination of effects due to the inherent stochastic nature of the switching critical current as well as other sources of variation such as noise and instrument error, forming the total standard deviation $\sigma_I$.

\subsection{\label{SuppE} Device location on chip}

\begin{figure}[h!]
\includegraphics[width=\columnwidth]{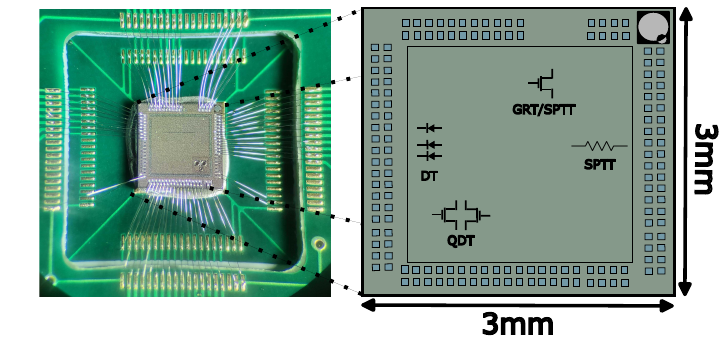}
\caption{\label{fig:supp_chip_locs}(Left) Picture of Bloomsbury chip, glued and wire-bonded to a PCB (Right) Diagram of Bloomsbury chip showing approximate locations of devices used for thermometry.}
\end{figure}

On the left hand side of Fig.~\ref{fig:supp_chip_locs}, we show an optical image of the chip used for these experiment. The chip is bonded to an FR4 PCB using Al:Si (1\%) bondwires. To the right, we show a schematic indicating the approximate location of the different thermometers and heaters.  

\subsection{\label{SuppF} Thermal Model}

\begin{figure}[ht!]
\includegraphics[width=\columnwidth]{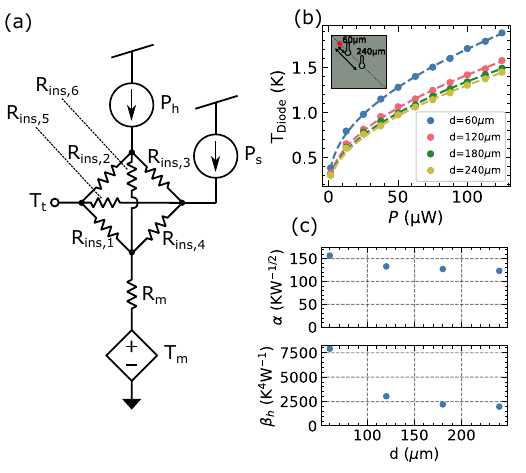}
\caption{\label{fig:5_Thermal_model} (a) Schematic of the thermal circuit used to model the system. (b) Diode temperature vs heater power for different separations, $d$. (c) Extracted parameters $\alpha$ and $\beta_\text{h}$ as a function of separation.}
\end{figure}

To fit the data in Fig.\ref{fig:4_remote_heating}, we developed a steady-state (no transient effects) thermal model including the temperature dependence of thermal conductivity of the various materials and the cooling power at the mixing chamber of the dilution refrigerator. The thermal model can be illustrated in the same way as an electrical circuit, in which power is represented by current and temperature is represented by voltage (see Fig. \ref{fig:5_Thermal_model}a). Working from the mixing chamber to the chip, the first component to model is the mixing chamber temperature. 
The mixing chamber cooling power is known to be proportional to the square of its temperature, $T_\text{m}$~\cite{Poole1983}. We can consider a power-balanced system in which the MXC will reach a stable temperature when its cooling power matches that of the combined background sources of power dissipation $P_\text{s}$ and any intentional on-chip heater power $P_\text{h}$ (assuming no other heat sources on the MXC). Thus, the temperature at the mixing chamber can be thought of as a power-controlled temperature source with the following functional dependence:

\begin{equation}
\label{eq:Tm}
    T_{\text{m}} = k\sqrt{P_{\text{h}}+P_{\text{s}}},
\end{equation}
where $k$ is a proportionality constant.

Next, we model the thermal conductivity, or equivalently thermal resistance, of the materials in the system. At low temperatures, the thermal conductivity of metals, such as Cu, Al and Au, scale linearly with temperature, while the thermal conductivity of insulators, such as Si, SiO$_2$ and Si$_3$N$_4$ in the chip, scale with the cube of temperature~\cite{Duthil2014}. The functional dependence of the thermal conductivity in metals is attributed to electronic thermal transport and in insulators to phononic thermal transport through the material lattice.
The MXC plate and the puck in which the PCB is placed are metallic (Au-plated OFC), meaning the thermal path from the MXC to the PCB can be modelled as one lumped thermal resistance $R_{\text{m}}$ that is inversely proportional to temperature. Furthermore, the electrical connections to the PCB are all metallic (PhBr). The net effective temperature value used for this proportionality is assumed to track linearly with the mixing chamber temperature (not necessarily equal to $T_{\text{m}}$, but some constant times $T_{\text{m}}$), meaning that the metal resistance, $R_{\text{m}}$, can be expressed as
\begin{equation}
\label{eq:Rm}
    R_{\text{m}} = a_{\text{m}}/T_{\text{m}},
\end{equation}
where $a_{\text{m}}$ is a proportionality constant.

Several metal contact points occur along the path represented by $R_{\text{m}}$, namely: mixing chamber plate to puck head, puck head to puck rails, and puck rails to PCB. The thermal resistance of the contact points is complex to model, as their thermal conductivities likely scale with temperature raised to the power of some unknown value in the range 0.75-2.5~\cite{salerno1997thermal}. For simplicity, we ignore the thermal contact resistance in this model and note that all contact points are Au-plated and were fastened as tightly and with as much surface area as possible to minimize the effects of thermal contact resistance in measurements. As we see in Fig.~\ref{fig:4_remote_heating}a, this approximation does not hinder the quality of the fit.

After the Au-plated contact between puck rail to PCB, the PCB and chip are both made of a combination of insulator and metal materials. However, we assume that the thermal transport in both cases is to be dominated by the phononic mechanism in the insulator portions which make up the bulk of the material.
Four distinct nodes must be modelled here representing four different physical locations in the system: the point at which the puck rail contacts the PCB, the location of the on-chip thermometer, the location of the on-chip heater, and the effective location of the on-chip background supply power dissipation. A complete mesh of thermal resistances between each of these nodes and each of the other three nodes is constructed, resulting in six thermal resistances $R_{\text{ins,1}}$ through $R_{\text{ins,6}}$. The temperature at the on-chip thermometer node is denoted as $T_{\text{t}}$. All six of these thermal resistances are modelled as proportional to the inverse of $T_{\text{t}}^3$ such that
\begin{equation}
\label{eq:Ri}
    R_{\text{ins},n} = a_{\text{ins,}n}/(T_{\text{t}}^3)
\end{equation}
where $n$ is a number between 1 and 6 and $a_{\text{ins},n}$ are constants. Again, this does not mean the effective temperature of each of the thermal resistances is exactly $T_{\text{t}}$, just that the effective temperature scales with $T_{\text{t}}$ in a constant relationship.
This assumption and the similar one for mixing chamber metal effective temperature are likely to be the most important sources of error of the model. One way to remove these assumptions would be to have a full 3D physical model of the system with temperature, heat flow, and thermal properties evaluated at many points along a spacial mesh. However, such an approach does not yield a useful analytical form of relationships like the model presented here.

The model is now sufficiently well-defined to determine $T_{\text{t}}$ as a function of $P_{\text{h}}$ and $P_{\text{s}}$. Tracing along the path from the mixing chamber to $T_{\text{t}}$ via $R_{\text{ins,1}}$ yields the following equation:
\begin{equation}
\label{eq:Tt}
    T_{\text{t}} = T_{\text{m}} + R_{\text{m}}(P_{\text{h}}+P_{\text{s}}) + R_{\text{ins,1}}(r_{\text{h}}P_{\text{h}}+r_{\text{s}}P_{\text{s}}),
\end{equation}
where $r_{\text{h}}$ is a value between 0 and 1 representing the portion of $P_{\text{h}}$ which flows through $R_{\text{ins,1}}$ and $r_{\text{s}}$ is a similar value representing the portion of $P_{\text{s}}$ which flows through $R_{\text{ins,1}}$. The values of $r_{\text{h}}$ and $r_{\text{s}}$ are constant and can be expressed in terms of $a_{\text{ins,1}}$ through $a_{\text{ins,6}}$. The key point is that the six resistances $R_{\text{ins,1}}$ through $R_{\text{ins,6}}$ form a constant power divider network through which $P_{\text{h}}$ and $P_{\text{s}}$ flow. Equation~\ref{eq:Tt} can then be rewritten as,
\begin{equation}
\label{eq:Tt4}
    T_{\text{t}}^4 = T_{\text{t}}^3\alpha\sqrt{P_{\text{h}}+P_{\text{s}}}+\beta_\text{h}P_{\text{h}}+\beta_\text{s}P_{\text{s}},
\end{equation}
where $\alpha=k+a_\text{m}/k$, $\beta_\text{h}=r_\text{h}a_\text{ins,1}$, and $\beta_\text{s}=r_\text{s}a_\text{ins,1}$. For a given $P_{\text{h}}$ and $P_{\text{s}}$, the positive real-valued solution for $T_{\text{t}}$ gives the modelled on-chip temperature. 

For each point in the measured data sets shown in Fig.~\ref{fig:4_remote_heating}, $T_{\text{t}}$, $P_{\text{h}}$ and $P_{\text{s}}$ are known. Thus, each data set can be be considered as a series of linear equations in $\alpha$, $\beta_\text{h}$, and $\beta_\text{s}$. Since each data set has more than three points, a Moore-Penrose inverse can be used to determine the least-squares solution for $\alpha$, $\beta_\text{h}$, and $\beta_\text{s}$. Each of these three values has a physical meaning useful for analysis. The parameter $\alpha$ represents the joint effects of metal thermal conductance in the system and cooling power at the mixing chamber changing with temperature. The parameters $\beta_\text{h}$ and $\beta_\text{s}$ represent the effect of thermal conductance of the insulating materials in the chip and PCB changing with temperature as they relate to the heater power and the background supply power, respectively. It is also worth noting that for more than two on-chip heat sources, this model can be generalized to include any number of heat sources as follows:
\begin{equation}
\label{eq:Tt4gen}
    T_{\text{t}}^4 = T_{\text{t}}^3\alpha\sqrt{\sum P_{\text{x}}}+\sum\beta_\text{x}P_{\text{x}}.
\end{equation}

Additionally, in Fig.~\ref{fig:5_Thermal_model}b, we plot the diode temperature versus heater power for different thermometer-heater separations. We fit the data to Eq.~\eqref{eq:Tt4}, considering $P_\text{s}\rightarrow 0$~W, to extract the thermal coefficients $\alpha$ and $\beta_\text{h}$ and plot them in Fig.~~\ref{fig:5_Thermal_model}c. We see a decay of the values as $d$ is increased consistent with the reduced heat transfer efficiency at larger distances.  

\begin{figure*}[ht!]
\includegraphics[width=\textwidth]{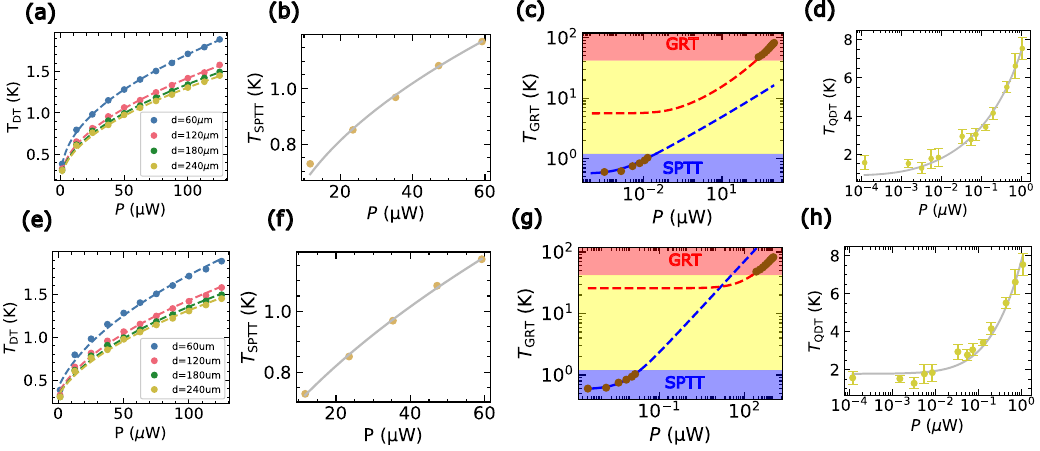}
\caption{\label{fig:10_Compare_Fits} Modelling of the heating data. (a)-(d) Model fits using Eq.~\eqref{eq:Tt4a} and (e)-(h) square root fits using Eq.~\eqref{eq:power}. For the data in (c) and (g), the blue fit extrapolates from the low-temperature (SPTT) points only, whereas the red fit extrapolates from the high-temperature (GRT) points only. As such, the red fit's low-temperature tapering in the yellow region should not be taken to represent a real effect.}
\end{figure*}

Finally, in Fig.~\ref{fig:10_Compare_Fits}, we show a comparison of quality of the fits to the heating data using the advanced model, including the effect of the insulating materials (Eq.~\eqref{eq:Tt4a}), and the simplified model (Eq.~\eqref{eq:power}); see panel a-d and e-h respectively. In Table 2, we compare the quality of the fits through $\chi^2$. We observe that the diode data benefits from including the contribution of the insulating materials, whereas for the rest of the thermometers (SPTT, GRT, and QDT) can be well fit by the square root depenence. 

\begin{table}
\label{Table2}
\centering
\begin{tabular}{|l|r|r|}
\hline
\textbf{Thermometer} & \textbf{$\chi^2$ full model} & \textbf{$\chi^2$ sqrt model} \\ \hline\hline
\textbf{\begin{tabular}[c]{@{}l@{}} Diode 60 $\mu$m \end{tabular}}  & 0.002 & 0.025 \\ \hline
\textbf{\begin{tabular}[c]{@{}l@{}} Diode 120 $\mu$m \end{tabular}}  & 0.009 & 0.018 \\ \hline
\textbf{\begin{tabular}[c]{@{}l@{}} Diode 180 $\mu$m \end{tabular}}  & 0.010 & 0.017 \\ \hline
\textbf{\begin{tabular}[c]{@{}l@{}} Diode 240 $\mu$m \end{tabular}}  & 0.014 & 0.016 \\ \hline
\textbf{\begin{tabular}[c]{@{}l@{}} SPTT \end{tabular}}  & 0.002 & 0.000 \\ \hline
\textbf{\begin{tabular}[c]{@{}l@{}} SGRT \end{tabular}}  & 0.013 & 0.001 \\ \hline
\textbf{\begin{tabular}[c]{@{}l@{}} QDT \end{tabular}}  & 0.749 & 0.622 \\ \hline
\end{tabular}
\caption{\label{Table2}$\chi^2$ of the fits in Fig.~\ref{fig:10_Compare_Fits} for the different sensor-heater arrangements. Middle column, fit to Eq.~\eqref{eq:Tt4a}. Right column) fit to Eq.~\eqref{eq:power}.}
\end{table}

\section{Acknowledgements}

All authors acknowledge Jonathan Warren and James Kirkman of Quantum Motion for their technical support during this work. T. H. S. acknowledges the Engineering and Physical Sciences Research Council (EPSRC) through the Centre for Doctoral Training in Delivering Quantum Technologies [EP/S021582/1]. M. F. G. Z. acknowledges a UKRI Future Leaders Fellowship [MR/V023284/1].

\section{Author Contributions}

G. M. N., M. K. and T. H. S. acquired the data. G. M. N., M. K., T. H. S. and M. F. G. Z. analysed the data. A. G. S. designed the IC. J. J. L. M. and M. F. G. Z. supervised the work. G. M. N. developed the thermal modelling. G. M. N., J. J. L. M. and M. F. G. Z. conceived the experiments. All authors contributed to the writing of this manuscript.  

\section{Data availability}

The data that support the plots within this paper and other findings of this study are available from the corresponding authors upon reasonable request.

\section{Competing interests}
The authors declare a relevant patent application: European Patent Application No. 23157647.1

\end{document}